\documentclass[aps,prb,floatfix,showpacs,twocolumn,footinbib]{revtex4}
\usepackage{epsfig}
\begin{document}


\title{Spin-polarized current amplification and spin injection in 
 magnetic bipolar transistors}

\author{Jaroslav Fabian} 
\affiliation{Institute for Theoretical Physics, Karl-Franzens University, 
Universit\"atsplatz 5, 8010 Graz, Austria}

\author{Igor \v{Z}uti\'{c}\footnote{Present address:
Center for Computational Materials Science,
Naval Research Laboratory, Washington, D.C. 20735, USA}}
\affiliation{Condensed Matter Theory Center, Department of Physics, 
University of Maryland at  College
Park, College Park, Maryland 20742-4111}

\begin{abstract}
The magnetic bipolar transistor (MBT) is a bipolar junction transistor with an equilibrium
and nonequilibrium spin (magnetization) in the emitter, base, or collector. The low-injection
theory of spin-polarized transport through MBTs and of a more general 
case of an array of magnetic {\it p-n} junctions is developed and illustrated on several 
important cases. Two main physical phenomena are discussed: electrical spin
injection and spin control of current amplification 
(magnetoamplification). It is shown
that a source spin can be injected from the emitter to the collector. If the 
base of an MBT has an equilibrium magnetization, the spin can be injected from the 
base to the collector by intrinsic spin injection. The
resulting spin accumulation in the collector is proportional to $\exp(qV_{be}/k_BT)$,
where $q$ is the proton charge, $V_{be}$ is the bias in the emitter-base
junction, and $k_B T$ is the thermal energy. 
To control the electrical current through MBTs both the equilibrium and
the nonequilibrium spin can be employed. The equilibrium spin controls the magnitude of the 
equilibrium electron and hole densities, thereby controlling the currents. 
Increasing the equilibrium spin polarization of the base (emitter) 
increases (decreases) the current amplification.   
If there is a nonequilibrium spin in the emitter, and the base or the emitter
has an equilibrium spin, a spin-valve effect can lead to a giant magnetoamplification
effect, where the current amplifications  
for the  parallel and antiparallel orientations of the 
the equilibrium and nonequilibrium spins differ significantly. 
The theory is elucidated using
qualitative analyses and is illustrated on an MBT example with generic materials parameters.
\end{abstract}
\maketitle

\section{Introduction}

Integrating  charge
and spin properties of semiconductors is the
central goal of semiconductor spintronics\cite{DasSarma2001:SSC} whose prospect 
has been fueled by the experimental demonstration of 
electrical spin injection into semiconductors, 
\cite{Osipov1998:PL,Hammar2002:PRL,Fiederling1999:N,Ohno1999:N,Jonker2000:PRB} 
as well as by the discovery of 
III-V ferromagnetic semiconductors\cite{Munekata1991:JCG,Ohno1998:S} 
(Eu-based ferromagnetic semiconductors have even been used
earlier as effective spin filters\cite{Esaki1967:PRL,Hao1990:PRB})
and observations of relatively long spin 
relaxation times.\cite{Kikkawa1998:PRL,Dzhioev2002:PRL} 
Many important advances have already been made toward an efficient
spin control of electrical current in semiconductors, and, vice versa, control
of magnetism by electrical means.  Recent examples include a control of ferromagnetism 
by incident light\cite{Koshihara1997:PRL,Oiwa2002:PRL} or 
by gate voltage,\cite{Ohno2000:N,Park2002:S} spin injection induced
magnetoresistance in nonmagnetic semiconductors,\cite{Schmidt2001:PRL} 
or the spin-galvanic effect.\cite{Ganichev2002:N}

Transistors are naturally suited for spin control of electrical currents
since the three regions, 
emitter, base, and collector, can serve as a
spin injector, transport medium, and spin detector, respectively.  
There has been remarkable experimental progress with employing
hybrid ferromagnetic metal and semiconductor structures as the
hot-electron transistors;\cite{Monsma1995:PRL} the magnetoresistance of such 
transistors can be as large as 3400\%\cite{VanDijken2003:PRL}
and they can be used as effective spin injectors\cite{Jiang2003:PRL}. 
The theoretical proposals for spin transistors focus largely on the
field-effect systems.\cite{Datta1990:APL,Ciuti2002:PRL,Schliemann2002:P} 
In this paper we analyze magnetic bipolar transistors (MBTs) which
are conventional (spin-unpolarized) 
bipolar junction transistors\cite{Shockley:1950}
with added spin.\footnote{The concept of a spin-polarized nonmagnetic bipolar junction transistor
was envisioned in the theoretical analysis of 
spin injection across conventional {\it p-n} junctions.\cite{Zutic2001:PRB}}  
MBTs were
first proposed in Ref.~\onlinecite{Fabian2002:P} (see also Ref.~\onlinecite{Fabian2003:P})
where iwe analyzed spin injection and current amplification of 
{\it npn} MBTs with a source spin.\footnote{The current paper
is an extended version of Ref.~\onlinecite{Fabian2002:P}.} 
Special cases of MBTs without a source spin were recently 
studied by Flatte et al.\cite{Flatte2003:APL} who calculated
the spin current polarization in a magnetic-base {\it npn} MBT, and
by Lebedeva and Kuivalainen\cite{Lebedeva2003:JAP}, who calculated the
current amplification in a magnetic emitter {\it pnp} MBT.  
Of spin transistors, the closest
one to MBT is the so called SPICE (spin polarized injection current
emitter),\cite{Gregg1997:JMMM} which employs ferromagnetic
metals in the emitter and base-collector regions. One of the principle
drives for proposing all-semiconductor spin transistors is the 
possibility of controlling current amplification  
by spin. 

MBTs integrate ferro(magnetic) and nonmagnetic semiconductors in the
usual bipolar junction transistor geometry.\cite{Shockley:1950,Tiwari:1992} Material 
and electrical properties of hybrid ferromagnet/semiconductor 
heterostructures are currently an active area of research.
\cite{Samarth2003:SSC} The potential of ferromagnetic semiconductors
for bipolar devices has been shown already in Ref.~\onlinecite{Wen1968:IEEETM}
where a ferromagnetic diode was presented.\footnote{We thank M. Field 
for bringing this reference to our attention.} 
More recently 
(Ga,Mn)As/GaAs {\it p-n} 
heterojunctions have been fabricated \cite{Ohno2000:ASS}
and electrical spin injection through magnetic bipolar tunnel junctions has
been demonstrated~\cite{Kohda2001:JJAP,Johnston-Halperin2002:PRB}
showing up to $\approx 80\%$ injected electron density spin polarization
at 4.6 K.\cite{vanDorpe2003:P}
Finally, in Ref.~\onlinecite{Tsui2003:APL} a CoMn doped p-Ge and an n-Ge 
were put together to
form a ferromagnetic {\it p-n} heterojunction which showed magnetization 
dependent current rectification, with up to 97\% electrical current variations 
due to the applied magnetic field. Such hybrid junctions
can also be used for MBTs, where the requirement is that the magnetic
region has a sizeable equilibrium spin polarization (say, 10\%). 
This polarization
can be provided by the exchange splitting in ferromagnetic semiconductors,
or by the large Zeeman splitting of dilute magnetic semiconductors. 
For example, Zeeman splitting can be significantly enhanced by
large effective $g$-factors in magnetically doped ($|g|\sim 500$ in Cd$_{0.95}$Mn$_{0.05}$Se 
at low temperatures)
or in narrow band gap semiconductors ($|g|\approx 50$ in InSb\cite{McCombe1971:PRB} even
at room temperature). Another possibility would be to use a
ferromagnetic semiconductor slightly above its Curie temperature,
a regime also expected to give large $g$-factors.
However, before there is an additional progress in fabricating junctions using
reported room temperature ferromagnetic semiconductors (for example,
(Zn,Cr)Te\cite{Saito2003:PRL}), the demonstration of the operation
of MBTs will likely be 
limited to temperatures below $\sim 150$ K.\footnote{Early prototypes of 
GaAs/(Ga,Mn)As-based  MBTs have recently been fabricated.
M. Field, private communication.}
Room temperature MBT is certainly a great challenge.
 
We formulate here 
a fully analytical theory of spin-polarized
transport through MBTs in the small bias (low injection)  regime, where the injected
carrier densities are smaller than the equilibrium ones. 
The theory uses the generalized Shockley model for the spin-polarized
transport through magnetic {\it p-n} junctions, \cite{Fabian2002:PRB}
as well as the theory of  conventional
bipolar junction transistors, as
developed by Shockley.\cite{Shockley:1950,Tiwari:1992}
Our theory can thus be viewed as a generalized Shockley theory of 
bipolar transistors.
Two novel phenomena are studied in detail: electrical spin 
injection from the emitter to the collector, and spin control
of the current amplification (also called gain). 
Electrical spin injection is shown to 
be effective in the amplification mode of the transistor, the
mode where the transistor amplifies current. Spin control
can be achieved by modifying both the equilibrium and nonequilibrium
spin, since both can modify the electrical current. The control
by the equilibrium spin (what we call the magnetoamplification effect)
results from the dependence of the equilibrium minority carrier
density on the equilibrium spin polarization, while the control by the
nonequilibrium spin (what we call the giant magnetoamplification effect) controls
the current via the spin-charge coupling of the Silsbee-Johnson
type.\cite{Silsbee1980:BMR,Johnson1985:PRL}

We first describe the model of MBTs in Sec.~\ref{sec:model} 
and formulate the analytical theory in Sec.~\ref{sec:theory}, 
leaving the formal aspects of the theory for the Appendix. We
then apply the theory to study electrical spin injection through 
MBTs in Sec.~\ref{sec:injection}, and spin control of the current 
amplification in Sec.~\ref{sec:electrical},
where we also discuss the spin current in MBTs.

\section{\label{sec:model} model}

\begin{figure}
\centerline{\psfig{file=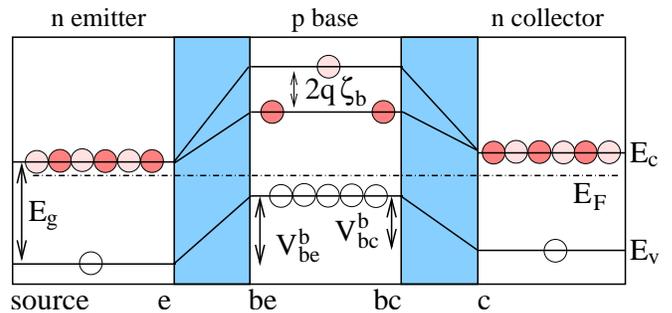,width=1\linewidth}}
\caption{Scheme of a magnetic {\it npn} transistor
in equilibrium.
The conduction band $E_c$ is populated mostly with electrons
(filled circles) in the
emitter and collector. In this example the base is magnetic
with the conduction band spin split by $2q\zeta_b$. The
valence band $E_v$, separated by the energy gap $E_g$ from
the conduction band is populated mostly by holes (empty circles)
in the base.
The small population of electrons in the base has an equilibrium
spin polarization $\alpha_{0b}=\tanh(q\zeta_b/k_BT)$, holes
are assumed spin unpolarized. The electron spin is indicated by
the dark (spin up) and light (spin down) shadings.
The Fermi level (chemical potential) $E_F$ is uniform. Between
the bulk regions, built-in potentials
$V_{be}^{b}$ and $V_{bc}^{b}$
are formed defining the depletion layers (shaded) in the
base-emitter ({\it b-e}) and base collector ({\it b-c}) junctions, respectively.
Finally, labels ``$\rm source$'',  $e$, $be$, $bc$, and $c$ stand for the
regions at which they are shown. For example, $be$ is the
region in the base at the boundary with the depletion layer.
}
\label{fig:scheme_eq}
\end{figure}

\begin{figure}
\centerline{\psfig{file=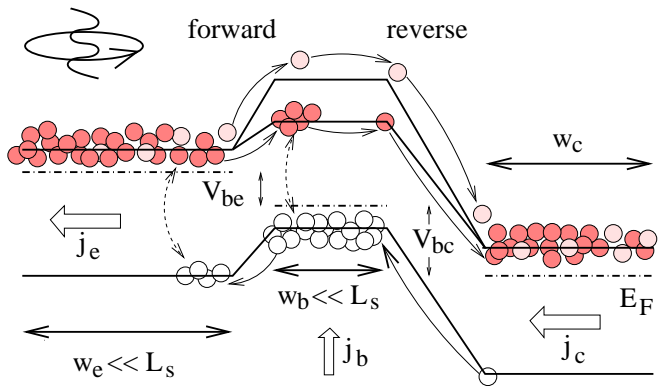,width=1\linewidth}}
\caption{Scheme of the magnetic {\it npn} transistor from Fig.~\ref{fig:scheme_eq}
in the amplification mode. The {\it b-e} junction is forward biased with $V_{be}>0$,
lowering the barrier and reducing the depletion layer width. The {\it b-c}
junction is reverse biased with $V_{bc}<0$, raising the barrier and
increasing the depletion layer width. The corresponding changes to the
Fermi  level
$E_F$ are indicated. The emitter has a spin source, indicated
here by the incident circularly polarized light generating nonequilibrium
electron spin well within the spin diffusion length $L_s$ from the {\it b-e}
depletion layer. The electron and hole flow 
gives  the emitter ($j_e$), base ($j_b$), and collector
($j_c$) charge currents. The electron-hole recombination is depicted
by the dashed lines. 
Also shown are the effective widths of the emitter
($w_e$), base ($w_b$), and collector ($w_c$). 
}
\label{fig:scheme_ne}
\end{figure}

A  conventional, spin-unpolarized
bipolar {\it npn} transistor \cite{Tiwari:1992} consists of $n$, $p$, and $n$ regions
connected in series (consult Figs.~\ref{fig:scheme_eq} and \ref{fig:scheme_ne}). 
Typically the $n$ region with the higher donor doping is
called the emitter, the one with the lower doping the collector. The base is
the $p$ region (doped with acceptors) sandwiched in between. 
The most useful mode of operation of the transistor is the so called
amplification (also forward active)
mode, where  the emitter-base ({\it b-e}) 
junction is forward biased, so that the electrons are easily injected into the base.
Together with the opposite flow of holes,
they form the emitter current $j_e$.
The electrons injected into the base diffuse towards the collector. 
The base-collector ({\it b-c}) junction is reverse biased. This means that any
electron reaching the junction from the base is swept by the electric
field in the depletion layer to the collector, forming the electron
current (holes's contribution is negligible).
The base current is the difference $j_b=j_c-j_e$.
This difference comes from two sources. First, from the hole current which
is present in the emitter but not in the collector. Second, from the electron-hole
recombination in the base which diminishes the number of electrons 
that make it to the collector. These two factors form the generally
small $j_b$. The current amplification 
$\beta$ is defined as the ratio of 
the large collector current to the small base current. 
For practical transistors $\beta$ is of order 100, meaning that  
small variations in $j_b$ (input signal)
lead to large variations in $j_c$ (output
signal). To maximize the  
gain one needs to (i) minimize
the relative contribution of holes in $j_e$, or (ii)  
inhibit the electron-hole recombination in the base. 
Typically silicon is used to make bipolar
transistors, since the indirect gap makes it a poor material for
the electron-hole recombination. We will show below that MBTs
allow spin control of the gain by realizing (i).

The magnetic bipolar transistor is a bipolar junction transistor with equilibrium
spin due to spin-split carrier bands, as well as with a nonequilibrium
source spin introduced, for example, by external electrical spin injection or optical
orientation. \cite{Meier:1984}
The equilibrium spin can be a result of 
the Zeeman splitting in an applied magnetic field or of the exchange splitting
due to ferromagnetic semiconductors integrated into the device structure.
For our purposes the equilibrium spin splitting should be on
the order of thermal energy 
for the spin-charge coupling discussed below to be significant. If no equilibrium
spin is present, this restriction becomes irrelevant, but the spin effects are limited
to electrical spin injection.

An MBT in equilibrium is described in Fig.~\ref{fig:scheme_eq}. The base is
magnetic, with the spin splitting $2q\zeta_b$. In the emitter the majority
carriers are electrons whose number is essentially $N_{de}$, the donor density.
Similarly in the collector, where the donor density is $N_{dc}$. Holes are the
minority carriers in the two regions. The base is doped with $N_{ab}$
acceptors. Holes (electrons) are the majority (minority) carriers.
We assume that only electrons are spin polarized. The inclusion
of the hole spin polarization is straightforward and adds no new physics
to our considerations. Furthermore, in many important semiconductors (such as GaAs) holes
lose their spin orientation very fast\cite{Hilton2002:PRL} 
and indeed can be treated as
unpolarized. Note that the electron density is $n=n_\uparrow+n_{\downarrow}$, 
the (electron) spin density is $s=n_\uparrow-n_{\downarrow}$, and the
spin polarization is $\alpha=s/n$. (Spin polarization of the 
charge current is studied in Sec.~\ref{sec:spin_current}.)

The equilibrium in MBTs can be disturbed
by applying a bias as well as
by introducing a nonequilibrium source spin. Figure \ref{fig:scheme_ne}
depicts the nonequilibrium physics and introduces the relevant notation. 
We assume  that the source spin is
injected into the emitter within the spin diffusion length from the 
{\it b-e} depletion layer so that enough spin can diffuse to the base.
At the {\it b-e} depletion layer the electrons feel a spin-dependent
barrier: in Fig.~\ref{fig:scheme_ne}
the barrier is smaller (larger) for the spin up (down) electrons.
As in  the conventional
bipolar transistors 
there is a significant accumulation of the minority carriers 
around the forward biased {\it b-e} depletion layer, 
while there are few carriers around the reverse biased {\it b-c} layer.
The widths {\it w} of the bulk regions depend on the applied
voltages as well as on the equilibrium spin polarization. \cite{Fabian2002:PRB}

We assume that the electron-hole recombination, occurring mostly 
in the emitter and the base is spin
independent, a reasonable approximation for unpolarized holes.
We also assume that the spin splitting is uniform in the bulk 
regions, eliminating magnetic drift 
(magnetic drift in semiconductors is discussed in Ref.~\onlinecite{Fabian2002:PRB}).
Our other assumptions are those of the standard Shockley theory. 
\cite{Shockley:1950,Tiwari:1992}: 
Temperature is large enough (say, $T\agt 50 K$) for all the donors and acceptors to be
ionized; The carriers obey the nondegenerate Boltzmann statistics;
The injected minority carrier densities are much smaller than the equilibrium
densities; The electric fields in the bulk regions are small eliminating
electrical drift. Furthermore, we neglect the carrier recombination and spin 
relaxation inside the depletion layers. 
These effects are important at 
very low biases and are not  
relevant for our observations, although may by themselves
lead to  nice physics. Finally, the contacts with the external electrodes
are ohmic, maintaining the carrier (but not necessarily spin) 
densities  in equilibrium.

\section{\label{sec:theory} Theory}

We generalize the Shockley theory of bipolar transistors to include
spin. The theory is valid in the small bias regime and is applicable
to any operational mode of the transistor, not only to the amplification
regime. Physically, the theory describes electron and hole carrier
and spin diffusion in the bulk regions, limited by the electron-hole
recombination and spin relaxation. The depletion layers provide only 
boundary conditions for the diffusion, by connecting the charge 
and spin currents in the adjacent regions. The most essential assumption
is that the spin-resolved chemical potentials remain constant across
the depletion layers.

The transistor is viewed as two {\it p-n} junctions in series.
The minority carrier density in each junction ({\it b-e} and {\it b-c})
is determined  by the bias voltage across each junction. In MBTs
the densities are determined also by the spin polarization, which needs
to be calculated self-consistently, as is explained below. Within the
limits of the theory it is enough to know the minority electron densities
$n_{be}$ and $n_{bc}$ to determine the electron charge currents, and
$p_e$ and $p_c$ to determine the hole charge currents (see Fig.~\ref{fig:scheme_eq}
for labeling the regions). We divide the presentation of the theory 
into two steps. First, we recall the main results of the generalized
Shockley theory of magnetic {\it p-n} junctions, \cite{Fabian2002:PRB}
and second, we use these results to formulate a theory of a series
of magnetic {\it p-n} junctions and solve it for {\it npn} MBT.
The first step is necessary to also understand
our qualitative analyses of the transistor operations in the amplification
mode. The second step, which is rather technical, is left for the Appendix.

\begin{figure}
\centerline{\psfig{file=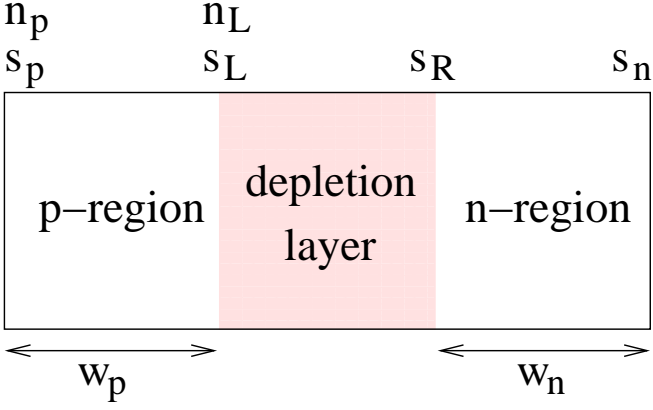,width=1\linewidth}}
\caption{Magnetic {\it p-n} junction. The input densities are
the carrier and spin densities at the end of the $p$ region:
$n_p$ and $s_p$, as well as at the end of the $n$ region: $s_n$. 
The densities to be calculated are at the
edge of the the depletion layer: 
$s_L$ in the $p$ and $s_R$ in the $n$ region. Also indicated
are the effective widths of the two bulk regions.  
}
\label{fig:pn}
\end{figure}

In the following we present selected results of the theory of
magnetic {\it p-n} junctions. The notation, which is easily
adapted for use in MBTs, is summarized in 
Fig.~\ref{fig:pn}. Both the $p$ and $n$ regions are in general
magnetic, biased with voltage $V$.
The nonohmic contact (to simulate the conditions at 
the base of a transistor) at the $p$ region maintains
nonequilibrium electron $n_p$ and spin $s_p$ densities.  Similarly,
there are nonequilibrium 
densities $n_L$ and $s_L$ at the 
left of the depletion layer. In the $n$ region electrons are
the majority carriers whose densities can be considered fixed
by the donor density $N_d$. However, the spin density can vary, 
being $s_R$ at the right of the depletion layer and $s_n$  at
the contact with the external ohmic electrode.  
We use subscript 0 to denote equilibrium quantities.
The equilibrium minority densities are $n_{0p}$ (electrons in $p$),
$p_{0n}$ (holes in $n$), and $s_{0p}$ (electron spin in $p$).
These densities are uniform across the corresponding regions.
The equilibrium density 
spin polarizations in the $n$ and $p$ regions
are $\alpha_{0n}$ and $\alpha_{0p}$, respectively. These are
also uniform. 
We denote the excess (injected) quantities
by $\delta$.  For example, 
$\delta s_L=s_L-s_{0p}$. We next denote by $L_{np}$ the
electron diffusion length in $p$, and by $L_{sn}$ and $L_{sp}$ 
the electron spin diffusion length in $n$ and $p$, respectively.
Finally, $D_{np}$ ($D_{nn}$) stand for the electron diffusion coefficients 
in $n$ ($p$). Similarly for holes.

The spin injection efficiency in magnetic {\it p-n} junctions
is measured by $\alpha_R=\delta s_R/N_d$, where\cite{Fabian2002:PRB}  
\begin{eqnarray} \label{eq:dsR}
\delta s_R &=& \gamma_0\delta s_n + \gamma_1 (\delta s_p-\alpha_{0p}\delta n_p) 
+\gamma_2 \alpha_{0L} \delta n_p \\
&-& \gamma_2 \cosh\left (w_p/L_{np}\right) s_{0L} \left (e^{qV/k_BT}-1\right );
\end{eqnarray}
the transport/geometry $\gamma$ factors are
\begin{eqnarray} \label{eq:gamma0}
\gamma_0&=&\frac{1}{\cosh(w_n/L_{sn})}, \\
\label{eq:gamma1}
\gamma_1&=&\left(\frac{D_{np}L_{sn}}{D_{nn}L_{sp}}\right)
\frac{\tanh(w_n/L_{sn})}{\sinh(w_p/L_{sp})},\\
\label{eq:gamma2}
\gamma_2&=&\left(\frac{D_{np}L_{sn}}{D_{nn}L_{np}}\right)
\frac{\tanh(w_n/L_{sn})}{\sinh(w_p/L_{np})}.
\end{eqnarray}
Equation (\ref{eq:dsR})
is accurate up to the terms of the relative order of $n_0\exp(qV/k_BT)/N_d$. While such terms
can be safely neglected when dealing with the spin and carrier densities, they must
be included when calculating the spin current in the $n$ region, where a difference
between two small spin densities of the same order needs to be evaluated. (These 
terms are not presented in Ref.~\onlinecite{Fabian2002:PRB}.) The exact formula
for the injected spin density $\delta s_R$ can be cast in the form of 
Eq.~(\ref{eq:dsR}), 
but with the coefficients $\gamma$ divided by the factor $1+\nu$: 
\begin{equation}\label{eq:gammap}
\gamma\rightarrow \gamma/(1+\nu),
\end{equation}
 where
\begin{eqnarray}
\nu&=&\frac{n_{0p}e^{qV/k_BT}}{N_d} [\gamma_1\cosh\left (w_p/L_{sp}\right )
\frac{1-\alpha_{0p}^2}{1-\alpha_{0n}^2} \\
&+&\gamma_3\alpha_{0p}\frac{\alpha_{0p}-\alpha_{0n}}{1-\alpha_{0n}^2}  ].
\end{eqnarray} 
Typically $\nu$ is a number smaller than 0.1, so the corrections to the 
spin injected density are upmost 10\%.
Knowing $\delta \alpha_R$ we can calculate the injected minority densities
$\delta n_L$ and $\delta s_L$:
\begin{eqnarray}\label{eq:dnL}
\delta n_L&=&n_{0p}\left [e^{qV/k_BT}\left(1+\delta \alpha_R \frac{\alpha_{0p}-
\alpha_{0n}}{1-\alpha_{0n}^2}  \right ) -1\right ],\\ \label{eq:dsL}
\delta s_L&=&s_{0p}\left [e^{qV/k_BT}\left(1+\frac{\delta \alpha_R}{\alpha_{0p}} 
\frac{1-\alpha_{0p}\alpha_{0n}}{1-\alpha_{0n}^2}  \right )  - 1\right ].
\end{eqnarray}
The following relation connects the spin polarization across the depletion layer:
\begin{equation} \label{eq:ap}
\alpha_L=\frac{\alpha_{0p}\left (1-\alpha_{0n}^2\right )+
\delta\alpha_R\left (1-\alpha_{0p}\alpha_{0n}\right )}
{1-\alpha_{0n}^2+\delta\alpha_R\left (\alpha_{0p}-\alpha_{0n}\right)}.
\end{equation}
Equations (\ref{eq:dsR}), (\ref{eq:dnL})-(\ref{eq:ap}) will be referred to as 
the {\it magnetic {\it p-n} junction equations}.

In the second step we wish to generalize the magnetic {\it p-n} junction
equations to the case of several magnetic {\it p-n} junctions in series.
Such a generalization is straightforward in the unpolarized case, where 
each junction acts independently from the others, since the minority
carrier densities are fixed only by $V$. The inclusion of spin complicates
the matter in the following sense. In a single junction 
$\delta n_p$, $\delta s_p$, and $\delta s_n$ are the known boundary conditions,
fully determining $\delta s_R$ and $\delta s_L$. Suppose we now connect
two junctions as in the {\it npn} MBT in Fig.~\ref{fig:scheme_eq}. Take the
{\it b-c} junction to be the one in Fig.~\ref{fig:pn}. Densities $n_p$
and $s_p$ become $n_{be}$ and $s_{be}$, themselves {\it unknown}, so that
$\delta s_R$ (now $\delta s_c$) is undetermined. On the other hand, considering
{\it b-e} to be the junction in Fig.~\ref{fig:pn}, $s_R$ becomes $s_e$ ($s_n$ becomes
the spin source density), 
and $n_L$ ($s_L$) become $n_{be}$ and $s_{be}$. These three densities
are determined also from $n_p$ and $s_p$, which are now 
denoted as $n_{bc}$
and $s_{bc}$. This loop shows the need to obtain the densities inside
the transistor (or a more general junction device) self-consistently. 
Charge and spin are coupled both across the depletion layers [through
Eqs.~(\ref{eq:dnL}) and (\ref{eq:dsL})]---intrajunction coupling---as well
as across the bulk regions between two depletion layers---interjunction
coupling. This theory is formally developed in the Appendix.

In the following we consider specific applications of the theory. Since
we will deal mostly with the amplification mode where the excess
densities in the {\it b-c} junction are negligible, we can get useful
insights even without the self-consistent solutions, using only the results
presented in this section. We refer to this as qualitative analysis. 
However, we support each case using a numerical example of a generic
MBT, calculated with the full theory presented in the Appendix. 
The numerical model is a ``silicon''-based MBT with the following 
room temperature parameters. 
Since the main features
of the full theory are captured by the qualitative formulas, one
can easily check the properties of MBTs with different parameters.
 The parameters
given below, while generic, are for illustration only.
Unless specified otherwise, the nominal widths of the emitter, base, and collector are 
$2$ $\mu$m, $1$ $\mu$m, and $2$ $\mu$m, respectively. The
donor doping densities of the emitter and collector are
$N_{de}=10^{17}$ cm$^{-3}$ and $N_{dc}=10^{15}$ cm$^{-3}$, while
the acceptor density in the base is $N_{ab}=10^{16}$ cm$^{-3}$. The
intrinsic carrier density at room temperature is $n_i=10^{10}$ cm$^{-3}$.
The carrier and spin relaxation times are taken to be 100 ns 
and 10 ns, and the electron (hole) diffusion coefficients 
$D_n=100$ ($D_p=10$) cm$^2$s$^{-1}$, all uniform throughout the
sample. The carrier and spin diffusion lengths are $L_{np}=(D_{np} \tau)^{0.5}\approx
30$ $\mu$m, $L_{pn}=(D_{pn} \tau)^{0.5}\approx 10$ $\mu$m, 
$L_{sn}=(D_{nn} T_1)^{0.5}\approx 10$ $\mu$m and $L_{sp}\approx L_{sn}$.
The dielectric constant is 12. We assume a spin ``ohmic'' contact
($\delta s=0$) at the end of the collector, while at the end 
of the emitter an external spin injection gives $\delta s^{\rm source}\ne 0$
in general.

\section{\label{sec:injection} Electrical spin injection}

Electrical spin injection through MBT 
will be studied in two cases: spin injection of the 
source spin from the emitter to the collector and spin injection
into the collector from the equilibrium spin in the base. The spin
injection efficiency in both cases is proportional to the Boltzmann
factor $\exp(qV_{be}/k_BT)$, but the physics behind them is rather
disparate. Unless specified otherwise, we work in the amplification
mode, where $V_{be} \gg k_B T$ (forward bias) and $V_{bc} \le 0$ (reverse bias),
and in the thin base limit, where $w_b \ll L_{nb}, L_{sb}$.

\subsection{Source spin}

Suppose a source spin density $\delta s^{\rm source}$ 
of polarization $\delta \alpha^{\rm source}=\delta s^{\rm source}/N_{de}$ is 
injected into the emitter. What is the spin response in the
collector? Consider first a nonmagnetic case ($\alpha_0=0$ everywhere).  
The spin injection involves three steps. 
(i) The source spin diffusion towards {\it b-e}. At the
depletion layer the nonequilibrium spin is 
$\delta s_{e}=\gamma_{0,be}\delta s^{\rm source}$. 
The spin polarization is $\delta\alpha_e=\delta s_{e}/N_{de}$. 
Note that $be$ in $\gamma_{i,be}$ ($\gamma_{i,bc}$) 
means that $\gamma_i$ given
in Eqs.~(\ref{eq:gamma0})-(\ref{eq:gamma2}) are evaluated for
the {\it b-e} ({\it b-c}) {\it p-n} junction. 
Since we assume that $L_{se}\agt w_e$, it follows that 
$\delta \alpha_{e}\approx \delta \alpha^{\rm source}$.
(ii) Transfer of the spin into the base. From Eq.~(\ref{eq:ap}) it
follows that $\alpha_e=\alpha_{be}$, showing the efficiency
of the spin injection by the majority electrons. 
The corresponding spin density is $\delta s_{be}=\delta \alpha_{be}
n_{0b}\exp(qV_{be}/k_B T)$, as follows from Eq.~(\ref{eq:dsL}) for the
forward bias case.
(iii) Spin injection into the collector. Equation (\ref{eq:dsR}) implies that
$\delta s_{c} = \gamma_{1,bc} s_{be}$ and so the nonequilibrium spin
polarization in the collector is $\alpha_{c}=\gamma_{1,bc} s_{be}/N_{dc}$,
a result of the minority electrons spin pumping:\cite{Zutic2001:PRB}
spin in the base diffuses towards
the reverse biased depletion layer where it is swept by the built-in field to the 
collector. Here the spin density accumulates as it is
bottle-necked by spin diffusion and spin relaxation.

The general formula for the spin injection, combining the processes (i) through (iii) 
in a magnetic transistor follows from the magnetic {\it p-n} junction equations:
\begin{equation} \label{eq:ac}
\delta \alpha_c=\delta \alpha^{\rm source} \gamma_{0,be}\gamma_{1,bc}
\frac{n_{0b}e^{qV_{be}/k_B T}}{N_{dc}}\frac{1-\alpha_{0b}\alpha_{0e}}{1-\alpha_{0e}^2}.
\end{equation}
In the small injection limit $n_{0b}\exp(qV_{be}/k_BT)/N_{dc}$ is small
(less than about 0.1); $\gamma_{0,be}$ is of order one. 
The spin injection efficiency increases with increasing $\gamma_{1,bc}$. 
In the thin base limit and for a wide collector ($w_c \gg L_{sc}$), for
example, obtains $\gamma_{1,bc}\approx L_{sc}/w_b$. This can be as large as
a ten or a hundred, making $\alpha_c$ a significant fraction of 
$\alpha_e\approx \alpha^{\rm source}$.
The decrease of $w_b$ can be achieved by increasing the width of the {\it b-c}
depletion layer, which, in turn, increases with increasing $|V_{bc}|$.

\begin{figure}
\centerline{\psfig{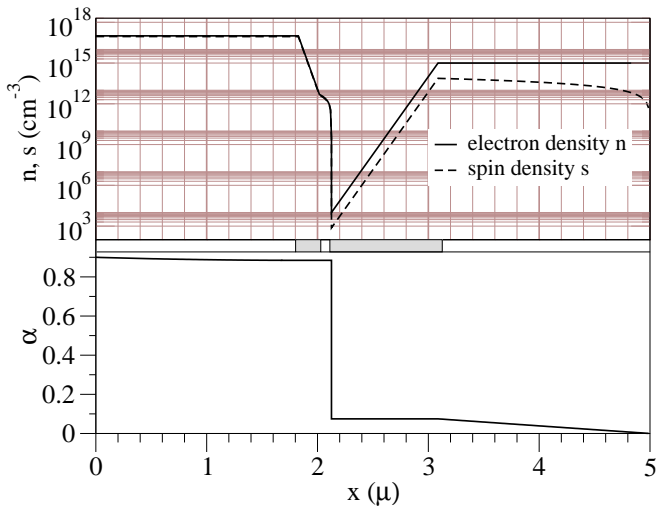}}
\caption{Calculated electron and spin density profiles (top) and
the spin polarization (bottom) in a nonmagnetic
{\it npn} transistor with a source spin of polarization 
$\delta \alpha^{\rm source}=0.9$
in the emitter.
The densities inside the depletion layers (shaded boxes) are not calculated;
they are shown, with no justification beside guiding the eye, as straight lines
connecting the densities at the depletion layer edges. 
Bias voltages are $V_{be}=0.5$ volt and $V_{bc}=0$ volt. 
Somewhat less than 10\% of the source spin polarization is 
transfered to the emitter.
}
\label{fig:2}
\end{figure}

Figure \ref{fig:2} illustrates the electrical spin injection of
the source spin in our numerical model system using the
full theory. The source spin of polarization $\delta \alpha^{\rm source}\approx 0.9$ 
first diffuses towards the base with a small decrease due
to spin relaxation. The spin polarization remains a constant
through the {\it b-e} depletion layer, resulting in a
nonequilibrium spin density in the base. The spin polarization
remains steady in the base, while both $n$ and $s$ decrease to their equilibrium values 
in going towards the {\it b-c} depletion layer. Right before the depletion
layer the spin polarization sharply drops, to get equal with
$\alpha_c \approx 0.075$. Such sharp drops are characteristic of the 
spin pumping by the minority carriers.\cite{Fabian2002:PRB}

\subsection{Equilibrium spin}

Is there a way to accumulate spin in an MBT without first injecting
a source spin into the structure? The answer is positive. In fact, there are two
different nonequilibrium spin densities accumulating as a result of
the carrier transport through the magnetic base. The first results from the
spin extraction, acting in the emitter, the second from the intrinsic spin injection,
effective in the collector. 

\begin{figure}
\centerline{\psfig{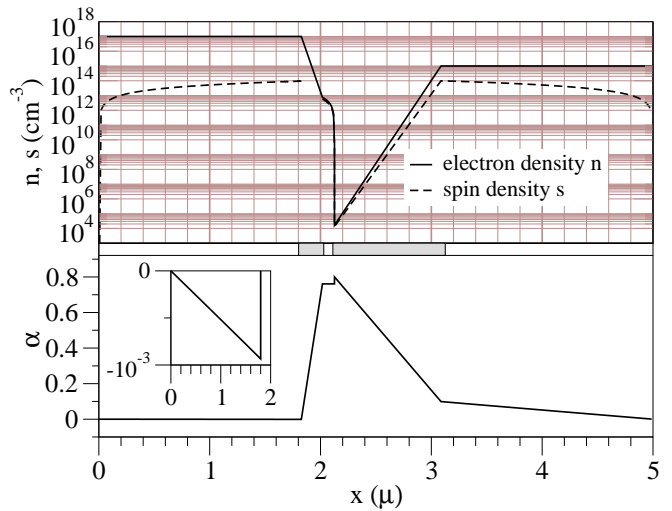}}
\caption{Spin extraction and intrinsic spin injection.
Calculated electron and spin density profiles (top) and
the spin polarization (bottom) in an {\it npn} transistor with a magnetic
base and nonmagnetic emitter and collector.
In the emitter region
spin density
$s<0$ (extraction), here plotted as positive ($s\rightarrow -s$);
the spin density is not shown in the {\it b-e} depletion layer.
The equilibrium spin polarization in the base is $\alpha_{0b}=0.762$,
corresponding to the conduction band splitting of 1 $k_B T$.
The bias voltages are $V_{be}=0.5$ volt and
$V_{bc}=0$ volt. The intrinsic spin injection, acting under the low
injection limit, results in the spin polarization in the collector
of $\delta\alpha_c\approx 10 \%$.
}
\label{fig:intrinsic}
\end{figure}

The extracted spin accumulates in a way similar to the magnetic diode.
\cite{Zutic2002:PRL,Fabian2002:PRB} The extracted spin density 
is small, on the order of the excess minority carrier densities. It is necessary
that the base has an equilibrium spin polarization. The emitter can, but
need not be, magnetic. Following Eq.~(\ref{eq:dsR}) we get
\begin{equation} 
\delta s_e=-\gamma_{2,be}\cosh(w_b/L_{nb})s_{0b}\exp(qV_{be}/k_BT).
\end{equation}
The result is a spin extraction from the emitter, since the
accumulated excess spin $\delta s_e$ has the opposite sign than the equilibrium spin
in the base.  The extracted
spin polarization $\delta \alpha_e=\delta s_e/N_{de}$
is small due to the generally large value of $N_{de}$.
This extracted spin density can also be treated as the spin source which
propagates to the emitter region, but the contribution 
to the collector spin is negligible, being of the order of
$[n_{0b}\exp(qV_{be}/k_BT)]^2/N_{de}N_{dc}$.

Intrinsic spin injection has no analog in the magnetic diode.  
The following physical processes are at work. (i) 
Minority electron injection into the base. 
The base has a spin-split conduction band, so the electrons
with the preferred spin will move at a faster rate, resulting 
in a {\it nonequilibrium} electron minority population, but with the
{\it equilibrium} spin polarization. The spin density is then out
of equilibrium. (ii) The nonequilibrium spin density at $be$ 
acts as a spin source in the {\it b-c} junction, similar to
the spin-polarized solar cell.\cite{Zutic2001:APL} 
(iii) This ``source'' spin is injected into the collector, where 
it accumulates. 

The result of the intrinsic spin injection, again in the limit
of the thin base, is
\begin{equation} \label{eq:ac2}
\delta \alpha_c= \alpha_{0b} \gamma_{0,be}\gamma_{1,bc}
\frac{n_{0b}e^{qV_{be}/k_B T}}{N_{dc}}.
\end{equation}
If both $\alpha^{\rm source}, \alpha_{0b}\ne 0$ 
the total spin polarization in the
collector is given by the sum of Eqs.~(\ref{eq:ac}) and (\ref{eq:ac2}). 
Remarkably, for $\alpha_{0e}=0$, the equilibrium spin polarization
$\alpha_{0b}$ in Eq.~(\ref{eq:ac2}) plays the role of $\delta\alpha_e$
in Eq.~(\ref{eq:ac2}). The equilibrium spin polarization behaves,
in MBTs, as a nonequilibrium source spin! This follows from
the spin-selective electrical injection across the {\it b-e} 
depletion layer.

Spin extraction and intrinsic electrical spin injection through 
an MBT are illustrated in Fig.~\ref{fig:intrinsic} using the full theory. 
The equilibrium polarization
spin in the base is
$\alpha_{0b}=0.762$. The electrical transport through the
base leads to a spin extraction from the emitter, with the
extracted spin polarization $\alpha_e \approx -0.001$, small
due to the large value of $N_{de}$. The spin polarization jumps to
its equilibrium value in the base, increasing sharply (see the discussion to
Fig.~\ref{fig:2}) to 
$\alpha_{0b}+\delta\alpha_c$ right before reaching the 
second depletion layer. The injected
spin polarization is $\delta \alpha_c\approx 10$\% , relatively large due to 
the small value of $N_{dc}$ and the large ratio $L_{sc}/w_b$ [see
Eq.~(\ref{eq:ac})].

We expect that both the direct injection of the source spin as well
as the spin extraction and the intrinsic spin injection become more
efficient in the limits of large carrier injection (large biases), where
our theory does not apply. This expectation is based on the results
of numerical calculations \cite{Zutic2002:PRL} of spin injection
in magnetic diodes. 

\section{\label{sec:electrical} electrical characteristics}

The electrical properties of MBTs are determined by both the charge
and the spin of the current carriers. There are
two ways spin affects the electrical currents: through the dependence of
the equilibrium minority electron and hole densities on the equilibrium spin polarization,
and through the spin-charge coupling resulting from the presence of a nonequilibrium
spin. We first introduce the formalism for calculating electrical currents 
in bipolar transistors and then analyze the two ways in detail. 
We conclude with a discussion of the spin current through MBTs.

\begin{figure}
\centerline{\psfig{file=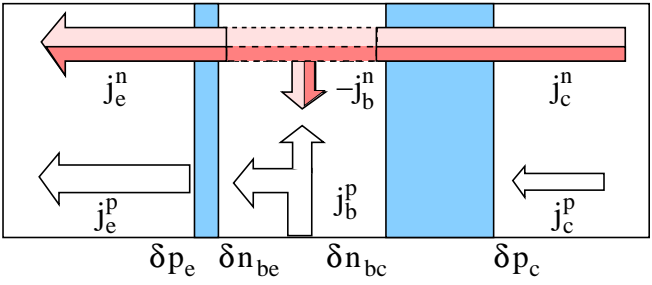,width=1\linewidth}}
\caption{Charge currents in an {\it npn} MBT. The electron emitter 
$j_e^n$ and collector $j_c^n$ currents are determined by the 
excess
electron densities $\delta n_{be}$ and $\delta n_{bc}$
in the base. Similarly, the hole emitter $j_e^p$ (collector $j_c^p$) 
current is determined by $\delta p_{e}$ ($\delta p_{c}$). 
The base current $j_b=j_e-j_c$ is formed by the electrons recombining 
with holes 
($j^n_b$) 
and by the holes that both recombine with 
electrons and enter the base from the base electrode ($j_b^p$).
Shading on the arrows of the electron currents indicate 
that the current is spin-polarized.
}
\label{fig:currents}
\end{figure}

The scheme and the sign convention for the currents 
is shown in Fig.~\ref{fig:currents} (see also Fig.~\ref{fig:scheme_eq}
for the description of symbols labeling the regions). Below we summarize
the expressions for the currents from the Shockley theory of
bipolar transistors.\cite{Shockley:1950,Tiwari:1992} We write the expressions in a rather
general form which turns out to be applicable also to MBTs (this
follows from the generalized Shockley theory of magnetic
diodes\cite{Fabian2002:PRB}). Let us define the generation 
current of a carrier $c$ (electron or hole) in region $r$
(emitter, base, or collector) as
\begin{equation} \label{eq:jg}
j_{gr}^c=\frac{qD_{cr}}{L_{cr}}c_{0r}\coth\left (\frac{w_r}{L_{cr}} \right ).
\end{equation}
The electron charge current density in the emitter is 
\begin{equation}\label{eq:jen}
j_e^n=j_{gb}^n\left [ \frac{\delta n_{be}}{n_{0b}}- 
\frac{1}{\cosh(w_b/L_{nb})} \frac{\delta n_{bc}}{n_{0b}}\right ].
\end{equation}
The first term in Eq.~(\ref{eq:jen}) represents the diffusion of electrons
in the base at the {\it b-e} junction. The second term describes
a competing diffusion from the excess minority electrons at the {\it b-e}
junction. Through the current continuity, this base diffusion current
continues to the emitter to become $j_e^n$. 
Similar expression holds for the electron current in the collector:
\begin{equation}
j_c^n=j_{gb}^n\left [-\frac{\delta n_{bc}}{n_{0b}}+
\frac{1}{\cosh(w_b/L_{nb})} \frac{\delta n_{be}}{n_{0b}}\right ].
\end{equation}
All the carrier densities
appearing in the expressions for the currents can be calculated from
the theory in Sec.~\ref{sec:theory}.
Holes contribute to the currents through the diffusion of their
excess minority populations $\delta p_e$ and $\delta p_c$. The
expression for the densities is similar to Eq.~(\ref{eq:dnL}) with 
no spin. For example,
\begin{equation}
\delta p_e= p_{0e} \left (e^{qV_{be}/k_B T} -1 \right ).
\end{equation}
The two hole currents are
\begin{eqnarray}
j_{e}^p&=&j_{ge}^p\frac{\delta p_{eb}}{p_{0e}}, \\
j_{c}^p&=&-j_{gc}^p\frac{\delta p_{cb}}{p_{0c}}.
\end{eqnarray}
The total emitter current is $j_e=j_e^n+j_e^p$, and similarly
the total collector current is $j_c=j_c^n+j_c^p$.
Finally, the base current is $j_b=j_e-j_c$. The task of computing the 
currents through an MBT is reduced to the computation of the 
excess electron and hole densities at the two depletion layers.  

The  current amplification coefficient (gain) $\beta$ is defined as
\begin{equation}
\beta=\frac{j_c}{j_b}.
\end{equation}
If $\beta$ is large, small changes in $j_b$ lead to large variations in $j_c$,
allowing signal amplification. The gain is often written as\cite{Tiwari:1992}  
\begin{equation} \label{eq:beta}
\beta=1/(\alpha_T' + \gamma').
\end{equation}
Here $\alpha_T'$ measures the (in)efficiency
of the electron-hole recombination in the base, and $\gamma'$ describes
the (in)efficiency of the emitter electron injection into the base. 
The usual base transport factor\cite{Tiwari:1992} 
is defined as $\alpha_T=1/(1+\alpha_T')$
and the emitter efficiency factor as $\gamma=1/(1+\gamma')$. 

After substituting for the currents, the emitter (in)efficiency $\alpha_T'$ 
is calculated to be
\begin{equation} \label{eq:at}
\alpha_T'= \cosh(\frac{w_b}{L_{nb}})-1.
\end{equation}
This is the value obtained for  conventional
transistors and is also valid
for MBTs. Coefficient $\alpha_T'$ does not depend on spin since it reflects
only the electron-hole recombination in the base and in our model the
recombination is spin independent. However, there may be 
cases where $L_{nb}$ depends significantly on $\alpha_{0b}$, in 
which case the gain could be controlled by spin
even through $\alpha_T'$. Note that $w_b$ depends on the equilibrium
spin through the spin dependence of the built-in fields.\cite{Fabian2002:PRB}
Eq.~(\ref{eq:at}) holds even for such cases.
On the other hand, $\gamma'$ depends explicitly on both the equilibrium and nonequilibrium
spin. We describe this dependence by defining a new parameter$\eta$:
\begin{equation}
\gamma'=\gamma'_0/\eta,
\end{equation}
where $\gamma'_0$ is the emitter efficiency of a conventional {\it npn} 
transistor\cite{Tiwari:1992} 
\begin{equation} \label{eq:g1}
\gamma_0'=\frac{N_{ab}D_{pe}n_{ie}^2}{N_{de}D_{nb}n_{ib}^2}\frac{L_{nb}\sinh(w_b/L_{nb})}
{L_{pe}\tanh(w_e/L_{pe})},
\end{equation}
where we allow for a generally different intrinsic carrier concentrations
$n_{ie}$ and $n_{ib}$ in the emitter and base, respectively.
In the next two sections we discuss the physics behind $\eta$, which we
call the magnetoamplification coefficient.
We will in particular consider the thin base limit, where
$\alpha_T'\sim (w_b/L_{nb})^2$, $\gamma'_0\sim (w_b/L_{nb})$, and $\gamma_0'$
dominates the current amplification   (for example in Si). 
In such cases 
\begin{equation} \label{eq:tb}
\beta \approx \eta/\gamma_0'. 
\end{equation}
If the base transport
factor is not negligible, the spin control efficiency diminishes.

\subsection{Magnetoamplification effect: influence of the equilibrium spin}

Consider a magnetic base. The Boltzmann statistics gives\cite{Zutic2002:PRL,Fabian2002:PRB}  
\begin{equation}
n_{0b}=\frac{n_{ib}^2}{N_{ab}}\tanh(q\zeta_b/k_B T)=
\frac{n_i^2}{N_{ab}}\frac{1}{\sqrt{1-\alpha_{0b}^2}}.
\end{equation}
Since $j_{gb}^n\sim n_{0b}$ [see Eq.~(\ref{eq:jg})], it follows that the 
base generation current increases as $\alpha_{0b}$ (that is, its magnitude) increases. In turn,
$j_e^n, j_c^n \sim j_{gb}^n$, so that both the emitter and the collector
currents increase with increasing $\alpha_{0b}$. The equilibrium spin
controls the charge currents flowing in MBT, leading to a magnetoresistance
effect. Spin unpolarized holes too contribute to the spin control of the
currents, as shown by Lebedeva and Kuivalainen\cite{Lebedeva2003:JAP}
for a {\it pnp} MBT.
If the emitter is magnetic, the minority hole density is
\begin{equation}
p_{0e}=\frac{n_{ie}^2}{N_{dc}}\frac{1}{\sqrt{1-\alpha_{0e}^2}}, 
\end{equation}
analogously for the collector (in the amplification mode
the hole density in the collector is negligible and does not affect
the current properties). The hole emitter current increases with increasing 
$\alpha_{0e}$.

\begin{figure}
\centerline{\psfig{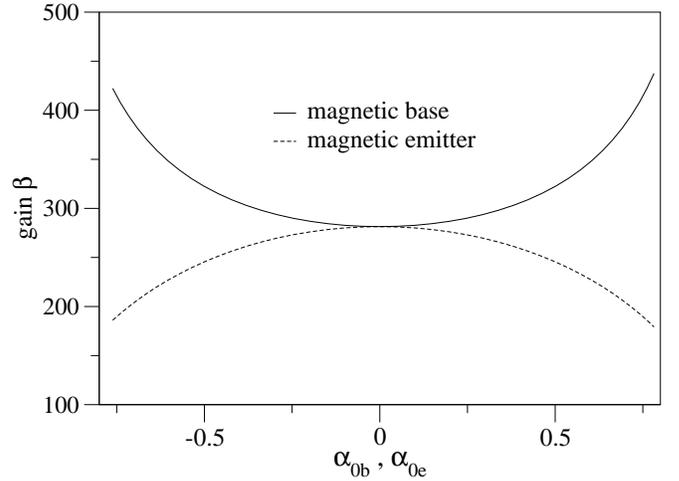}}
\caption{Calculated gain $\beta$ of an MBT 
with a magnetic base (solid) and emitter (dashed) for our numerical model. 
No source spin is present. The equilibrium base (emitter) spin polarization
is $\alpha_{0b}$ ($\alpha_{0e}$). The calculation is done on a structure with
the nominal base width of 1.5 $\mu$m to keep the effective width
$w_b$ positive for the considered range of polarizations.  
}
\label{fig:gaineq}
\end{figure}

The magnetoamplification coefficient amounts to the simple expression
\begin{equation}
\eta=\sqrt{\frac{1-\alpha_{0e}^2}{1-\alpha_{0b}^2}}, 
\end{equation}
and in the thin base limit, Eq.~(\ref{eq:tb})
is thus described by the  
gain
\begin{equation} \label{eq:MR}
\beta = \frac{1}{\gamma_0'}\sqrt{\frac{1-\alpha_{0e}^2}{1-\alpha_{0b}^2}}.
\end{equation}
The gain of MBTs can thus be  controlled by controlling
the equilibrium magnetization (for example, by changing the external
magnetic field) of the emitter or the base. 
The collector magnetization plays no role. On the other hand, the 
two equilibrium polarizations $\alpha_{0e}$ and $\alpha_{0b}$ 
act against each other: The  
gain increases (decreases) with a greater
spin splitting in the base (emitter). This is because the emitter (in)efficiency
$\gamma'$ increases (decreases) if there are relatively more holes (electrons) present 
in $j_e$. If the spin polarization is uniform across the {\it b-e} 
junction, the gain is spin independent.

The opposite role of the equilibrium magnetizations in the base and in the
emitter is shown in Fig.~\ref{fig:gaineq}, which illustrates the behavior
of $\beta$ with respect to the changes of $\alpha_{0b}$ and $\alpha_{0e}$
separately. The calculation is done using the full theory, not the
approximate formulas above. However, the approximation describes
the calculation very well, showing that spin dependent effects for
example on the effective widths $w$, which are accounted for in
the full theory, play minor role in our example.

\subsection{Giant magnetoamplification effect: spin-charge coupling in MBT}

A nontrivial realization of the Silsbee-Johnson spin-charge 
coupling,\cite{Silsbee1980:BMR,Johnson1985:PRL} representing the
physics of the proximity of an equilibrium and nonequilibrium
spin in MBTs is what we call here the giant magnetoamplification
effect (GMA), in analogy with giant magnetoresistance 
(GMR) effect in metallic multilayers.\cite{Maekawa:2002b} 
For GMA it is necessary that there be a nonequilibrium
spin polarization in the emitter (arising from a source spin) and
an equilibrium spin either in the base or in the emitter (or both). 
The physics is illustrated in Fig.~\ref{fig:scheme_ne}. The charge current through MBTs 
depends on the relative orientation of the source and the
equilibrium spins, because of the spin-dependent barrier
in the {\it b-e} junction.

In the presence of a nonequilibrium spin density $\delta \alpha_e$, the
magnetoamplification  coefficient 
becomes
\begin{equation} \label{eq:sigma}
\eta=\sqrt{\frac{1-\alpha_{0e}^2}{1-\alpha_{0b}^2}}\left [1+\delta\alpha_e
(\alpha_{0b}-\alpha_{0e})/(1-\alpha_{0e})^2 \right ].
\end{equation}
If only the base is magnetic, the gain in the thin base limit is 
\begin{equation} \label{eq:bgma}
\beta = \frac{1}{\gamma_0'}\frac{1+\delta\alpha_e \alpha_{0b}}{\sqrt{1-\alpha_{0b}^2}}.
\end{equation}
The spin-charge coupling is described by the product $\delta \alpha_e \alpha_{0b}$, 
similar to implications of the spin-voltaic effect in magnetic {\it p-n}
junctions.\cite{Zutic2002:PRL,Fabian2002:PRB,Zutic2003:P}
Let $\beta_{\rm max}$ and $\beta_{\rm min}$ are the  
gains for
the configuration of the source and equilibrium spins (parallel or
antiparallel) that yield the maximum and minimum gain, respectively.
For a magnetic base (emitter) the maximum is achieved at parallel (antiparallel) 
orientation and the minimum at antiparallel (parallel) orientation of the 
source and equilibrium spins, respectively.
We define the GMA coefficient as
\begin{equation} \label{eq:gma}
GMA=\frac{\beta_{\rm max}-\beta_{\rm min}}{\beta_{\rm min}},
\end{equation} 
in analogy with a similar expression (involving resistivities) defining
the GMR coefficient.

For the magnetic base
\begin{equation}
GMA=\frac{2|\delta\alpha_e\alpha_{0b}|}{1-|\delta\alpha_e\alpha_{0b}|}.
\end{equation}
If, for example, $\delta\alpha_e=\alpha_{0b}=0.5$, GMA=67\%. 
The analogy with GMR is clear: there is a large magnetoresistance effect (greater than
10\%), which is most pronounced when the relative orientation
of two spin polarizations changes from parallel to antiparallel. 
If, on the other hand, the emitter is magnetic, the effect is opposite: the parallel
spin orientation decreases the gain, due to the decrease in the
the emitter injection efficiency. The GMA coefficient is
\begin{equation}
GMA=\frac{2|\delta\alpha_e\alpha_{0e}|}{1-\alpha_{0e}^2+|\delta\alpha_e\alpha_{0e}|}.
\end{equation}
The GMA coefficient vanishes if $\alpha_{0e}=\alpha_{0b}$. 
To decide on whether to use a magnetic base or
a magnetic emitter one needs to take into account that a magnetic
base will have a smaller $L_{sb}$. If $L_{sb} \alt w_b$, a magnetic
emitter would be instead preferable. Note that the GMA coefficient is directly
proportional to the magnitude of $\delta \alpha_e$, and so it can
be used to measure the nonequilibrium spin polarization, as in the 
case of magnetic diodes.\cite{Zutic2003:APL}

\begin{figure}
\centerline{\psfig{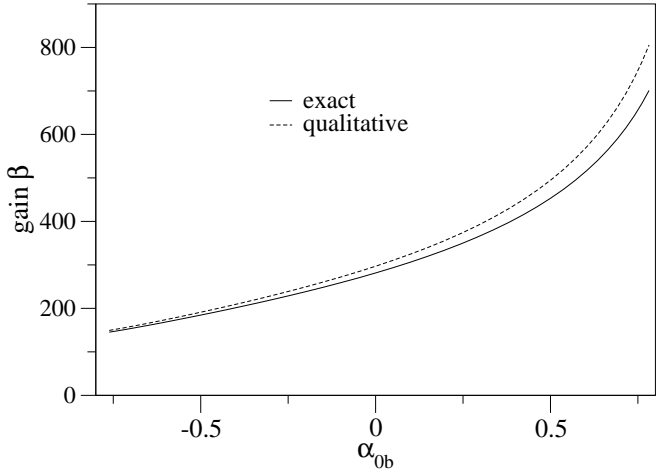}}
\caption{Giant magnetoamplification effect. 
Calculated gain $\beta$ of our MBT example with a magnetic base
and source emitter spin, as a function of $\alpha_{0b}$ for
a fixed $\alpha^{\rm source}=0.9$. The solid curve is the
calculation using the full theory, the dashed curve is
the contribution of the emitter efficiency only, $\beta \approx
\eta/\gamma'$, where the magnetoamplification 
coefficient
$\eta$
is given by Eq.~(\ref{eq:sigma}).
The calculations are done on a structure with the base
long 1.5 $\mu$m, as in Fig.~\ref{fig:gaineq}
}
\label{fig:gainne}
\end{figure}

Figure \ref{fig:gainne} illustrates GMA
on our MBT example with a magnetic base  and source spin polarization
$\alpha^{\rm source}=0.9$ (as in Fig.~\ref{fig:scheme_ne}). The solid line
represents the calculation of $\beta(\alpha_{0b})$ 
using the full theory, while the dashed line
is the approximation Eq.~(\ref{eq:bgma}) valid for the thin base transistors.  
The approximative formula works very well. 
The asymmetric curve demonstrates the GMA effect. When the equilibrium
and the source spin are antiparallel, $\beta$ is small;
when they are parallel, $\beta$ is large. The magnetic control of the
charge current and its amplification is thus predicted to be rather
effective.

\subsection{\label{sec:spin_current} Spin-polarized current}

Thus far we have studied the spin polarization of the electron density
as a measure of the spin injection efficiency. This spin polarization is typically 
detected by optical experiments. Another 
spin polarization, that of the charge current, is more convenient
for theory and is invariably
used in spin injection models  to assess the spin injection efficiency
(see the $\gamma$-technique of Rashba\cite{Rashba2000:PRB,Rashba2002:EPJ})
and to establish the boundary conditions for spin at the interfaces.\cite{Fabian2002:PRB}
The current spin polarization $\alpha_j$ is the ratio of the spin 
current $j_s$, which is the difference between the charge currents formed
by the spin up and spin down electrons, and the total charge current $j$:
$\alpha_j=j_s/j$. 
The current spin polarization is much less intuitive than 
the density spin polarization. There are several reasons for that.
First, $\alpha_j$ involves not only the carrier and spin densities, 
but also the drift or diffusion velocities which can be spin dependent. 
Next, unlike $\alpha$, the magnitude of $\alpha_j$ is not restricted to the interval 
$(0,1)$. The magnitude of $\alpha_j$ is not 
even bound. Indeed, the spin current can be finite even if there 
is no charge current, making $\alpha_j$ infinite.
Also the 
signs of $\alpha_j$ and $\alpha$ can be different. Finally, unlike
the charge current, the spin current need not be uniform.
Because of spin relaxation, $j_s$ (and also $\alpha_j$) 
is not conserved. For the above reasons, unless the relation
between $\alpha$ and $\alpha_j$ is obvious or is 
explicitly derived, $\alpha_j$ is not indicative of the spin injection
efficiency. In particular in inhomogeneous (or hybrid) semiconductors
at degenerate
doping densities or at large biases, both 
spin diffusion and spin drift are relevant and one needs to 
employ the Poisson equation to solve the transport problem 
self-consistently to obtain a dependence between $\alpha$ and
$\alpha_j$. Many of the experimental spin injection results
are likely to fall in this category, making realistic theoretical
modeling difficult. 

Fortunately, in the low injection limit there is a simple relation
between the spin current and the spin, so the knowledge of $\alpha_j$
together with the knowledge of the charge current $j$ suffice to 
obtain $\alpha$. 
The spin current density at point $c$ (see Fig.~\ref{fig:scheme_eq})
is readily obtained from Tab.~II in Ref.~\onlinecite{Fabian2002:PRB}:
\begin{equation}
j_{sc}=\frac{qD_{nc}}{L_{sc}}\coth\left (w_c/L_{sc}\right) \delta s_c.
\end{equation}
Thus, $\delta \alpha_c=\delta s_c/N_{dc}$ is directly proportional to
$j_{sc}=j \alpha_j$. As is shown below, $\alpha_j$ is usually comparable
to 
$\alpha_0$ or $\delta \alpha$, 
largely independent on the biases. The spin injection
efficiency is then determined by $j_c$, which, in turn, depends
exponentially on $V_{be}$. 

We adopt the same sign convention for the spin currents as for the charge
currents, see Fig.~\ref{fig:currents}. 
A straightforward application of the magnetic {\it p-n} junction equations 
(Sec.~\ref{sec:theory})
gives  
\begin{equation} \label{eq:ajc}
\alpha_{jc}=\frac{e^{qV_{be}/k_BT}\left (\alpha_{0b} 
+\delta\alpha_{e}\frac{1-\alpha_{0b}\alpha_{0e}}{1-\alpha_{0e}^2}\right ) -\alpha_{0b} }
{e^{qV_{be}/k_BT} 
\left (1+\delta\alpha_{e}\frac{\alpha_{0b}-\alpha_{0e}}{1-\alpha_{0e}^2} \right ) -1}
\end{equation}
What is striking (although not so surprising) 
is that $\alpha_{jc}$ in the cases of practical biases $|V_{be}|\gg k_B T$
is bias independent. Indeed, for a forward bias $V_{be}$, that is, in the 
amplification mode, 
\begin{equation} \label{eq:ajc1}
\alpha_{jc}=\frac{\alpha_{0b}(1-\alpha_{0e}^2)+\delta \alpha_e (1-\alpha_{0b}\alpha_{0e})}
{1-\alpha_{0e}^2+\delta \alpha_e (\alpha_{0b}-\alpha_{0e})}.
\end{equation}
This is in sharp contrast, for example, to  Eq.~(\ref{eq:ac}) which 
displays the exponential increase of the spin injection
efficiency with $V_{be}$. 
In the limit of no source spin, Eq.~(\ref{eq:ajc1}) reduces to $\alpha_{jc}=\alpha_{0b}$,
the main result of Flatte et al.\cite{Flatte2003:APL} Similarly, if
the transistor has no equilibrium spin, the spin current polarization is
$\alpha_{jc}=\delta \alpha_{e}$. If the emitter/base bias is reverse, $V_{be} < 0$,
spin injection is practically nonexisting. Yet, there is a large current 
spin polarization, $\alpha_{jc}=\alpha_{0b}$, independent of the source 
spin, confirming our claim that large $\alpha_j$ alone does not imply 
efficient spin injection.

\begin{figure}
\centerline{\psfig{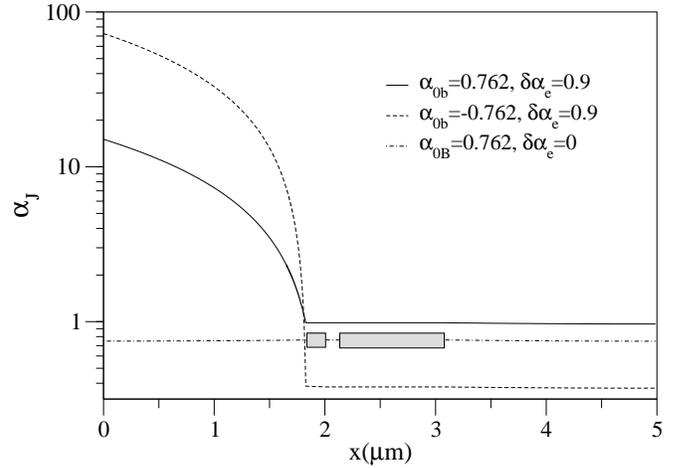}}
\caption{Calculated spin current polarization profile in the magnetic {\it npn} transistor,
calculating by the full theory.
The source spin at the emitter is fixed at $\delta \alpha^{\rm source}=0.9$, 
while the equilibrium base spin polarization changes sign at the magnitude
of $\alpha_{0b}=0.762$, corresponding to the base spin splitting of 
the order of the thermal energy. The biases are $V_{be}=0.5$ volt and $V_{bc}=0$ volt.
The spin current remains uniform across the depletion layer, based on the 
assumption of negligible spin relaxation there. The spin current is normalized
by the emitter and the collector currents in the emitter and the collector
regions, and by the base electron current in the base. The horizontal line
is for no source spin. The spin current stays uniform throughout the sample at the
value given by $\alpha_{0b}$, implying no significant spin relaxation, but not
necessarily spin injection.}
\label{fig:spin-current}
\end{figure}

Our analysis is illustrated in Fig.~\ref{fig:spin-current}.
We consider the amplification mode, but the results vary
little with the applied bias, as explained above.  Figure 
\ref{fig:spin-current} describes three cases. One with the source 
spin and the equilibrium spin pointing in the same direction, 
one where the two spins are antiparallel, and one where only the equilibrium
spin is present. In all the cases $\alpha_{jc}$ (which is the value
at $x\approx 3$ $\mu$m) is smaller than 1, but rather considerable,
consistent with Eq.~(\ref{eq:ajc1}). Note that $\alpha_j$ in 
the emitter is much greater than 1 for the case of the source spin,
due to the spin diffusion current being greater than the majority
electron drift current there. The value of $\alpha_j$ decreases 
at the emitter/base junction, where it becomes a constant, signifying
small spin relaxation in both the base and in the emitter. Finally,
the case of no source spin polarization ($\delta \alpha_e \ll \alpha_{0b}$)
shows a flat $\alpha_{j}$. This case, viewed from the density spin
polarization perspective, is shown in Fig.~\ref{fig:intrinsic}. The nonequilibrium spin
which accumulates in the base is injected to the emitter, and extracted
from the collector. The injection/extraction efficiency increases
exponentially with $V_{be}$. For small biases, even if $\alpha_{j}$ is
equal to $\alpha_{0b}\approx 0.762$, the spin injection is negligible. 
Interestingly, even if the spin current polarization is positive in
the emitter, the spin density polarization is negative (spin extraction,
cf. Fig.~\ref{fig:intrinsic}).

\section{Conclusions}

We have developed an analytic theory for the spin-polarized transport in magnetic
bipolar transistors in the low injection regime. We have shown that the
transistor displays a number of novel phenomena, not observed either in
the conventional spin unpolarized bipolar transistor or in the magnetic diode. One
such effect is the intrinsic electrical spin injection, which is a spin
injection from a magnetic base into a nonmagnetic collector, without 
the presence of a source spin in the emitter. The spin injection efficiency
depends exponentially on the emitter/base bias. 
Other effects are related to the  
gain which can be influenced in two ways.
First, by the equilibrium spin, either in the emitter or in the base, 
modifying the generation current. Second, by the spin-charge coupling,
modifying directly the electron injection from the emitter. 

Bipolar junction transistors are used for ultra high speed logic applications thanks
to the fast carrier transport. Spin can bring additional functionalities.
The magnetoamplification effects can be used to study spin signals (time varying
spin polarizations) by detecting charge currents. Indeed, all the currents
in an MBT in the amplification mode depend primarily on $\delta n_{be}$,
which is controlled by both the equilibrium and nonequilibrium spin. 
The spin-to-current conversion can thus be observed by measuring the collector
current, or directly the GMA coefficient Eq.~(\ref{eq:gma}). On the other hand, the amount of the
injected spin in the collector depends exponentially on $V_{be}$ in the same manner
as the charge current depends on it. Changes in, for example,  
the base current, can thus lead to changes in the spin polarization $\delta \alpha_c$. 
As a result, current signals can be detected by observing the accumulated
spin polarization. Perhaps the most attractive feature of the magnetoamplification
effects is that the spin splitting or the source spin polarization are
not built-in device properties but can change on demand, during the 
transistor operation, by magnetic field. This is why an MBT is an example of a variable
heterostructure transistor.  

Another use of MBTs may be in the electrically induced magnetization switching,
similarly to what has been observed as light-induced ferromagnetism
\cite{Koshihara1997:PRL,Oiwa2002:PRL} or ferromagnetism induced by the gate voltage
of field-effect transistors. \cite{Ohno2000:N,Park2002:S}
If the base is a ferromagnetic semiconductor, the equilibrium 
magnetization depends on the density of free carriers. This density can be,
in turn, controlled by $V_{be}$. While the scenario of the electrically
induced ferromagnetism in a nondegenerate MBT is probably not realistic,
at higher doping and current injection levels (where our theory no longer applies) 
this effect could be observable. 

We believe that the phenomena we propose to study are robust and should
be observed. For the spin-source spin injection one does not need a magnetic
semiconductor in the structure. The source spin can be generated in
the emitter either optically or electrically, and similarly the spin injected
into the collector can be observed by detecting 
the polarization of electroluminescence.\cite{Fiederling1999:N,Jonker2000:PRB} 
For the phenomena related
to the equilibrium magnetism, the rule of thumb is that the spin splitting
should be comparable to the thermal energy in order to create noticeable spin 
polarizations. While to see GMA at room temperature may be difficult at present, 
at smaller temperatures (starting from 50 K where the shallow donors and acceptors
start to ionize) the effects could be detected at the spin splitting levels of 
$\sim 1$ meV.  

\acknowledgments{We are grateful to S. Das Sarma for support and encouragement,
and for useful discussions. I. \v{Z}. acknowledges 
the National Research Council for financial support. 
This work was funded by DARPA, the NSF-ECS, and the US ONR. 
}

\appendix

\section{\label{appendix} An array of magnetic  {\it p-n} junctions}

We introduce a formalism for evaluating the carrier and spin
densities in an array of magnetic {\it p-n} junctions. An array of two junctions
forms an MBT, while three junctions would form a magnetic thyristor. The boundary
conditions are applied for the densities; it is straightforward to adapt
the method to have boundary conditions specified by the spin currents.
All the junctions need to be considered simultaneously since there is
both {\it intra-} and {\it inter-} junction charge and spin coupling. 
The intrajunction coupling arises from the generalized Shockley 
conditions\cite{Fabian2002:PRB} of the uniformity of the spin-resolved
chemical potentials and of the continuity of the charge and spin currents,
across a junction's depletion layer. The interjunction
coupling arises from the carrier and spin diffusion in the
bulk regions between the depletion layers: the carrier and spin density
at one end influences the current at the other end, and vice versa. The two couplings
lead to a set of linear algebraic equations for the densities.

\begin{figure}
\centerline{\psfig{file=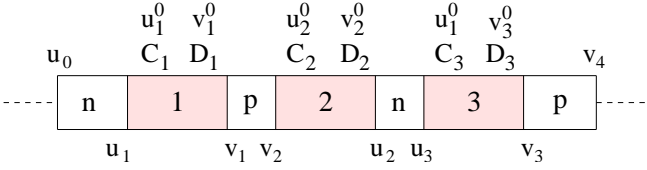,width=1\linewidth}}
\caption{Notation scheme for an array of {\it p-n} junctions. The figure
illustrates an {\it npnp} (thyristor) structure. The depletion layers are
shaded. The known quantities are shown above the junctions, the unknown
below.  Each junction $i$ is characterized by the scalar $u_i^0$ and by the
triplet of vectors
${\bf v}^0_i$, ${\bf C}_i$, and ${\bf D}_i$, which are determined  by
the doping densities and the equilibrium carrier and spin densities
of the regions adjacent to the junction, and by the applied voltage
$V_i$ across the junction.  
The nonequilibrium spin density $u_0$ (note that the symbol $0$
here denotes the region, not an equilibrium) and the
charge/spin density vector ${\bf v}_4$ would be the boundary conditions here.
The densities $u_i$ and ${\bf v}_i$ at the depletion
regions are to be obtained self-consistently. These
densities are all what is needed to calculate the charge and spin currents.
}
\label{fig:algebra}
\end{figure}

The array we consider is shown in Fig.~\ref{fig:algebra}.
Each junction is given a number $i$ starting from one. The junctions are
either of the {\it p-n} ($p$ left and $n$ right), 
or of the {\it n-p} ($n$ left and $p$ right) type. In the following the
indexes $n$ and $p$ relate to the $n$ and $p$ regions adjacent to
the junction in question. Let us introduce the notation using
a generic junction, as in Sec.~\ref{sec:theory}.
The bias $V$ across the junction is positive
for the forward and negative for the reverse direction of the charge current.
The electron density is $n$, with index zero ($n_0$) reserved for the equilibrium
value. The nonequilibrium (excess) part of the density is denoted
as $\delta n = n-n_0$. 
Similarly for the electron spin density $s$ and for the spin polarization
$\alpha=s/n$. The doping densities are $N_d$ for the donors in the $n$ region
and $N_a$ for the acceptors in the $p$ region.

We make the complex notation more compact by introducing some unifying symbols. 
We first denote by $u^0$ the scalar characterizing the nonequilibrium
spin density due to the carrier extraction:
\begin{equation}
u^0=-\gamma_2 \cosh(w_p/L_{np}) s_{0p}\left (e^{qV/k_B T}-1\right).
\end{equation}
Here $k_BT$ is the thermal energy, with $T$ denoting temperature; $q$ is the
proton charge. Parameters $\gamma$ are introduced in 
Eqs.~(\ref{eq:gamma0})-(\ref{eq:gamma2}). To properly account for the spin current
in the $n$ region, one needs to apply the rescaling in Eq.~(\ref{eq:gammap}).
We also introduce vector ${\bf v}^0$ which specifies the
nonequilibrium electron and spin density in the $p$ region
as a result of the carrier injection:
\begin{equation}
{\bf v}^0=\left (e^{qV/k_B T}-1\right)\left [n_{0p},s_{0p}\right].
\end{equation}
The dimensionless vector $\bf C$ characterizes intrajunction 
coupling:
\begin{equation}
{\bf C}=\left [\alpha_{0p}(\gamma_2-\gamma_1),\gamma_1\right],
\end{equation}
while another dimensionless vector $\bf D$, given by 
\begin{equation}
{\bf D}=\frac{n_{0p}}{N_d} \frac{e^{qV/k_B T}}{1-\alpha^2_{0n}}
\left [\alpha_{0p}-\alpha_{0n},1-\alpha_{0p}\alpha_{0n}\right ],
\end{equation}
characterizes interjunction coupling. The quantities
$u^0$, $\bf v^0$, $\bf C$, and $\bf D$ are presumed to be
known. They are the input parameters.

The unknown quantities are the nonequilibrium carrier and spin densities at the 
depletion layers. In the low-bias regime considered here, the
electron density in the $n$ regions is fixed: $n_n\approx N_d$, and only $\delta s_n$ 
needs to be calculated. In the $p$ regions the excess electron (spin)
density $\delta n_p$ ($\delta s_p$) is unknown. 
We denote by scalar $u$ the nonequilibrium spin density $\delta s_n$ 
in the adjacent $n$ region:
\begin{equation}
u=\delta s_n.
\end{equation}
Vector ${\bf v}$ will describe both the nonequilibrium electron and spin
density in the {\it p} region:
\begin{equation}
{\bf v} = \left[\delta n_p, \delta s_p \right ].
\end{equation}
The boundary conditions are given by the corresponding $u$ or $\bf v$ (depending
on whether the $n$ or the $p$ region is the contact region) at the
beginning and the end of the array.

Using the above notation, the magnetic  {\it p-n} junction equations 
Eqs.~(\ref{eq:dsR}), (\ref{eq:dnL}), and (\ref{eq:dsL})
are greatly simplified and can be adapted to solve the array problem. 
Indeed, for junction $i$ the {\it p-n} junction  equations can be written as
\begin{eqnarray} \label{eq:i}
u_i&=&u_i^0+\gamma_{0,i} u_{i\pm 1} +{\bf C}_i\cdot {\bf v}_{i\mp 1}, \\
{\bf v}_i& =& {\bf v}_i^0 + {\bf D}_i u_i,
\end{eqnarray}
where the upper (lower) sign is for the {\it p-n} ({\it n-p}) directed junction.
For the {\it npn} MBT described in the main text, the 
equations take the form
\begin{eqnarray} \label{eq:u1}
u_1&=&u_1^0+\gamma_{0,1} u_0 +{\bf C}_1\cdot {\bf v}_2, \\ \label{eq:v1}
{\bf v}_1& =& {\bf v}_1^0 + {\bf D}_1 u_1,
\end{eqnarray}
for junction $1$, and
\begin{eqnarray} \label{eq:u2}
u_2&=&u_2^0\gamma_{0,2} u_3 +{\bf C}_2\cdot {\bf v}_1, \\ \label{eq:v2}
{\bf v}_2& =& {\bf v}_2^0 + {\bf D}_2 u_2,
\end{eqnarray}
for junction 2. Analogous equations can be written for holes. 
The solution to Eqs.~(\ref{eq:u1})-(\ref{eq:v2}) is
\begin{equation} \label{eq:solution}
u_2=\left ({\bf C}_2\cdot{\bf D}_1 \right )\left (\gamma_{0,1}u_0+u_1^0 \right)
+ {\bf C}_2\cdot {\bf v}^0_1 + u_2^0+ \gamma_{0,2} u_3,
\end{equation}
where we have neglected the terms of order $[n_{0p}\exp(V)/N_d]^2$, small in
the low bias regime. The formulas for $u_1$, ${\bf v}_1$, and ${\bf v}_2$ can be
obtained directly by substituting Eq.~(\ref{eq:solution}) back to
Eqs.~(\ref{eq:u1}) through (\ref{eq:v2}).

Equation (\ref{eq:solution}) describes spin injection through an MBT, since $u_2$ is
the nonequilibrium spin in the collector at the depletion layer with the base. 
The first term on the right-hand side (RHS) of Eq.~(\ref{eq:solution})
represents the transfer of source spin $u_0$
from the emitter to the collector. Indeed, for a nonmagnetic transistor (the equilibrium
spin polarizations are zero) the
transferred source spin is $u_3=\gamma_{0,1}\gamma_{1,2}n_{0p}\exp(V_1/k_BT)u_0$. Here
$\gamma_0$ describes the transfer of the source spin through the emitter---a majority
carrier spin injection. Once the spin is in the base, it becomes the spin of the
minority carriers [hence the minority density factor 
$n_{0b}\exp(V_1/k_BT)$], 
diffusing towards the base/collector depletion layer 2. The built-in electric field in
this layer sweeps the spin into the collector, where it becomes the spin of
the majority carriers again, by the process of the minority-carrier spin
pumping.~\cite{Zutic2001:PRB,Fabian2002:PRB}
Can the
injected spin polarization in the collector be greater than the source spin polarization?
The answer is negative in the low-injection regime. It would be tempting to let the
spin diffusion length in the collector to increase to large values to get
a greater pumped spin. But that would increase the importance of the electric
field in the $n$-regions and the theory (which is based on the charge and spin diffusion and
not on the spin drift) would cease to be valid. However, the spin density in the collector
can be greater than that in the base,
demonstrating that the spatial decay of the nonequilibrium spin (spin accumulation)
is not, in general, monotonically
decreasing, in line of what was demonstrated nonmagnetic {\it p-n}
junctions\cite{Zutic2001:PRB}  as well as inhomogeneous unipolar semiconductors.\cite{Pershin2003:PRL}
The same term also describes the transfer of the nonequilibrium
spin $u_1^0$ accumulated as a result of the electron injection into the magnetic base.
The second term of Eq.~(\ref{eq:solution}),  which is independent of the source spin,
results from the (intrinsic) spin pumping by the minority
channel of the {\it nonequilibrium} spin generated in the base by the forward current
through junction 1. This term vanishes if the base is nonmagnetic ($\alpha_{0b}=0$).
The third term of Eq.~(\ref{eq:solution})
represents the spin extraction due to the magnetic base. This term
is controlled by $V_2$, the bias at the base/collector junction. For the 
reverse bias, used in the amplification mode, this term can be neglected.
Finally, the last term of Eq.~(\ref{eq:solution}), describes the diffusion
of the source spin in the collector.

The knowledge of the carrier and spin densities at the depletion layers 
allows us to calculate the charge and spin currents in the systems, as
well as the density spatial profiles throughout the bulk regions, using
the formulas for magnetic {\it p-n} junctions in Tab.~II of Ref.~\onlinecite{Fabian2002:PRB}.
This is done for our numerical MBT model presented in Secs.~\ref{sec:injection}
and \ref{sec:electrical}.

\bibliographystyle{apsrev}
\bibliography{references}

\end{document}